\documentclass[twocolumn,superscriptaddress,showpacs,preprintnumbers,prl,amsmath,amssymb,
final]{revtex4}  
\usepackage{graphicx}
\usepackage{units}
\usepackage{xspace}

\def\EeV{\ifmmode {\mathrm{\ Ee\kern -0.1em V}}\else
                   \textrm{Ee\kern -0.1em V}\fi}%
\def\eV{\ifmmode {\mathrm{\ e\kern -0.1em V}}\else
                   \textrm{e\kern -0.1em V}\fi}%
\def\gcm{\ifmmode {\mathrm{g/cm}^2}\else
                   {g/cm$^2$}\fi\xspace}%
\def\Xmax{\ifmmode {X_\mathrm{max}}\else
                   {$X_\mathrm{max}$}\fi\xspace}%
\def\sigmaXmax{\ifmmode {\mathrm{RMS}(X_\mathrm{max})}\else
                   {RMS$(X_\mathrm{max})$}\fi\xspace}%
\def\meanXmax{\ifmmode {\langle X_\mathrm{max}\rangle}\else
                   {$\langle X_\mathrm{max}\rangle$}\fi\xspace}%
\newcommand{\depth}[1]{\unit[#1]{\gcm}}
\newcommand{\energy}[1]{\unit[$10^{#1}$]{\eV}}
\newcommand{\minimumEnergy}{\energy{18}\xspace}
\newcommand{\numberOfEvents}{3754\xspace}
\newcommand{\approxNumberOfEvents}{four thousand\xspace}
\newcommand{\firstData}{December 2004\xspace}
\newcommand{\lastData}{March 2009\xspace}
\newcommand{\lowD}{\unit[(106$^{+35}_{-21}$)]{\gcm/decade}\xspace}
\newcommand{\highD}{\unit[(24$\pm$3)]{\gcm/decade\xspace}}
\newcommand{\breakE}{\energy{18.24\pm 0.05}\xspace}
\newcommand{\highRMS}{55}
\newcommand{\lowRMS}{26}

\begin{document}

\title{
  Measurement of the Depth of Maximum of Extensive Air Showers above
  \unit[$\mathbf{10^{18}}$]{\eV}
}
\date{\today}
\begin{abstract}
  We describe the measurement of the depth of maximum, \Xmax, of the
  longitudinal development of air showers induced by cosmic rays. 
  Almost \approxNumberOfEvents events above \minimumEnergy
  observed by the fluorescence detector of the Pierre Auger
  Observatory in coincidence with at least one surface detector
  station are selected for the analysis. The average shower maximum was
  found to evolve with energy at a rate of \lowD below \breakE and
  \highD above this energy. The measured shower-to-shower fluctuations
  decrease from about \highRMS~to \depth{\lowRMS}. The interpretation
  of these results in terms of the cosmic ray mass composition is briefly
  discussed.
\end{abstract}
\collaboration{The Pierre Auger Collaboration}
\noaffiliation
\author{J.~Abraham}
\affiliation{National Technological University, Faculty Mendoza (CONICET/CNEA), Mendoza, Argentina}
\author{P.~Abreu}
\affiliation{LIP and Instituto Superior T\'{e}cnico, Lisboa, Portugal}
\author{M.~Aglietta}
\affiliation{Istituto di Fisica dello Spazio Interplanetario (INAF),  Universit\`{a} di Torino and Sezione INFN, Torino, Italy}
\author{E.J.~Ahn}
\affiliation{Fermilab, Batavia, IL, USA}
\author{D.~Allard}
\affiliation{Laboratoire AstroParticule et Cosmologie (APC), Universit\'{e} Paris 7, CNRS-IN2P3, Paris, France}
\author{I.~Allekotte}
\affiliation{Centro At\'{o}mico Bariloche and Instituto Balseiro (CNEA-UNCuyo-CONICET), San Carlos de Bariloche, Argentina}
\author{J.~Allen}
\affiliation{New York University, New York, NY, USA}
\author{J.~Alvarez-Mu\~{n}iz}
\affiliation{Universidad de Santiago de Compostela, Spain}
\author{M.~Ambrosio}
\affiliation{Universit\`{a} di Napoli ``Federico II'' and Sezione INFN, Napoli, Italy}
\author{L.~Anchordoqui}
\affiliation{University of Wisconsin, Milwaukee, WI, USA}
% ...

\maketitle

{\it Introduction} -- The energy dependence of the mass composition of
cosmic rays is, along with the flux and arrival direction
distribution, an important parameter for the understanding of
the sources and propagation of cosmic rays at very high energy. There
are several models that describe the observed flux of cosmic
rays very well, but each of these models has
different assumptions about the cosmic ray sources and correspondingly
predicts a different mass composition at Earth. For example,
the hardening of the cosmic ray energy spectrum at energies between
\energy{18} and \energy{19}, known as the 'ankle', is presumed to be
either a signature of the transition from galactic to extragalactic
cosmic rays or a distortion of a proton-dominated extragalactic
spectrum due to energy
losses~\cite{ankle}. Moreover, composition
information may eventually help to decide whether the flux suppression
observed above~4$\cdot$\energy{19}~\cite{bib:gzkmeas} is due mainly to
the interaction of cosmic rays with the microwave background or a
signature of the maximum injection energy of the
sources~\cite{Allard:2008gj}.

Due to the low flux at these energies, the composition of cosmic rays
cannot be measured directly, but has to be inferred from observations
of extensive air showers.  The atmospheric depth, \Xmax, at which the
longitudinal development of a shower reaches its maximum in terms of
the number of secondary particles is correlated with the mass of
the incident cosmic ray particle.
With the generalization of Heitler's model of electron-photon cascades
to hadron-induced showers and the superposition assumption for nuclear
primaries of mass $A$, the average depth of the shower maximum,
\meanXmax, at a given energy $E$ is expected to
follow~\cite{bib:heitlerModel}
\begin{equation}
   \meanXmax = \alpha\left(\ln E-\langle\ln A\rangle\right) + \beta,
   \label{eq:XmaxSuperposition})
\end{equation}
where $\langle\ln A\rangle$ is the average of the logarithm of the primary
masses.  The coefficients $\alpha$ and $\beta$ depend on the nature of
hadronic interactions, most notably on the multiplicity, elasticity
and cross-section in ultra-high energy collisions of hadrons with air,
see e.g.\ \cite{Ulrich:2009hm}. 
Although
Eq.~(\ref{eq:XmaxSuperposition}) is based on a simplified description of
air showers, it gives a good description of air shower
simulations with energy-independent parameters $\alpha$ and $\beta$ in
the energy range considered here,
see~\cite{Pierog:2006qu}.  Only physics processes not
accounted for in currently available interaction models
could lead to a significant energy dependence of these parameters.

The change of \meanXmax per decade of
energy is called {\itshape elongation rate}~\cite{bib:elongationRate},
\begin{equation}
   D_{10}=\frac{\mathrm{d}\meanXmax}{\rm{d} \lg E} \approx \alpha
     \left(1-\frac{\mathrm{d}\langle\ln A\rangle}{\rm{d} \ln E}\right)\ln(10),
\end{equation}
and it is sensitive to changes in composition with energy.
A complementary composition-dependent observable 
is the magnitude of the shower-to-shower fluctuations
of the depth of maximum, \sigmaXmax,
which is expected to decrease with the number of primary
nucleons $A$ (though not as fast as $1/\sqrt{A}$
\cite{bib:fluctuations}) and to increase with the interaction
length of the primary particle.

At ultra high energies, the shower maximum can be observed directly
with fluorescence detectors. Previously published \Xmax
measurements~\cite{Bird:1993yi,bib:Abbasi2005ni} focused mainly on
\meanXmax as a function of energy and had only limited statistics
above \energy{19}. 

Here we present a measurement of both \meanXmax and
\sigmaXmax using high quality and high statistics data collected with
the southern site of the Pierre Auger
Observatory~\cite{bib:auger}. The Observatory is located in the province of
Mendoza, Argentina and consists of two detectors. The surface detector
(SD) array comprises 1600 water-Cherenkov detectors arranged on a
triangular grid with \unit[1500]{m} spacing that cover an area of over
\unit[3000]{km$^2$}.  The water-Cherenkov detectors are sensitive to
the air shower components at ground level.  The fluorescence detector
(FD) consists of 24 optical telescopes overlooking the array, which
can observe the longitudinal shower development by detecting the
fluorescence and Cherenkov light produced by charged particles along
the shower trajectory in the atmosphere.

{\it Data Analysis.} -- This work is based on air shower data recorded
between \firstData and \lastData.  Only events detected in hybrid
mode~\cite{bib:hybrid} are considered, i.e.\ the shower development
must have been measured by the FD, and at least one coincident SD
station is required to provide a ground-level time. Using the time
constraint from the SD, the shower geometry can be determined with an
angular uncertainty of 0.6$^\circ$~\cite{bib:angReso}.  The
longitudinal profile of the energy deposit is
reconstructed~\cite{bib:profileRec} from the light recorded by the FD
using the fluorescence and Cherenkov yields and lateral distributions
from~\cite{bib:lightyields}. With the help of data from atmospheric
monitoring devices~\cite{bib:augeratmo} the light collected by the
telescopes is corrected for the attenuation between the shower and the
detector and the longitudinal shower profile is reconstructed as a
function of atmospheric depth. \Xmax is determined by fitting the
reconstructed longitudinal profile with a Gaisser-Hillas
function~\cite{bib:gaisser-hillas}.

\begin{figure}[!t]
\includegraphics[clip,bb=29 0 575 384,width=\linewidth]{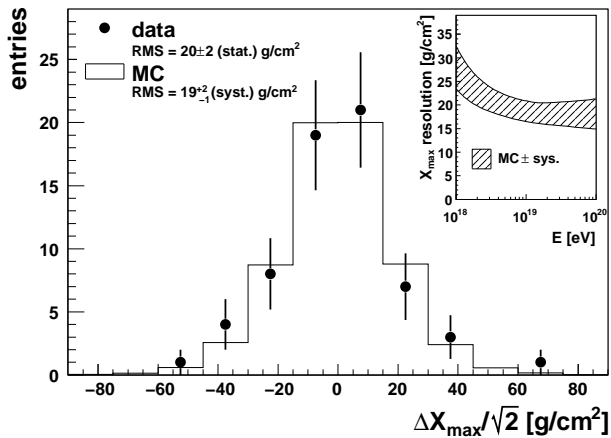}
\caption{Difference between \Xmax measured in showers simultaneously at
  two FD stations ($\langle\lg(E/\!\!\eV)\rangle=19.1$). The \Xmax
  resolution is displayed as a function of energy in the inset.}
\label{fig_resolution}
\end{figure}
An unbiased set of high quality events is selected 
with the statistical uncertainty of the
reconstructed \Xmax being comparable to the size of the fluctuations
expected for nuclei as heavy as iron ($\approx$ \depth{20}) and small
systematic uncertainties as explained in the following.

The impact of varying atmospheric conditions on the \Xmax measurement
is minimized by rejecting time periods with cloud
coverage and by requiring reliable measurements of the
vertical optical depth of aerosols. Profiles that are distorted by
residual cloud contamination are rejected by a loose cut on the
quality of the profile fit ($\chi^2$/Ndf$<$2.5).
We take into account events only with energies above \energy{18} where
the probability for at least one triggered SD station is 100\%,
irrespective of the mass of the primary particle~\cite{bib:icrcflux}.
The geometrical reconstruction of showers with a large apparent
angular speed of the image in the telescope is susceptible to
uncertainties in the time synchronization between FD and SD.
Therefore, events with a light emission angle towards the FD that is
smaller than 20$^\circ$ are rejected.  This cut also removes events
with a large fraction of Cherenkov light.
The energy and shower maximum can be reliably measured only if \Xmax
is in the field of view (FOV) of the telescopes (covering
$1.5^\circ$ to $30^\circ$ in elevation). Events for which only the
rising or falling edge of the profile is detected are not used.
Moreover, we calculate the expected statistical uncertainty of the
reconstruction of \Xmax for each event, based on the shower geometry
and atmospheric conditions, and require it to be better than \depth{40}.

The latter two selection criteria may cause a selection bias due to
a systematic undersampling of the tails of the true \Xmax
distribution, since showers developing very deep or shallow in the
atmosphere might be rejected from the data sample. To avoid such a 
bias in the measured \meanXmax and \sigmaXmax we apply
fiducial volume cuts based on the shower geometry that ensure that the
viewable \Xmax range for each shower is large enough to accommodate the
full \Xmax distribution~\cite{bib:fovcuts}.

After all cuts, \numberOfEvents events are selected for the \Xmax
analysis. The \Xmax resolution as a function of energy for these
events is estimated using a detailed simulation of the FD and the
atmosphere. As shown in the inset of Fig.~\ref{fig_resolution}, the
resolution is at the \depth{20} level above a few \EeV.  The
difference between the reconstructed \Xmax values in events that had a
sufficiently high energy to be detected independently by two or more FD
stations is used to cross-check these findings. As can be seen in
Fig.~\ref{fig_resolution}, the simulations reproduce the data well.
\begin{figure}[!t]
\includegraphics[clip,bb=29 0 575 384,width=\linewidth]{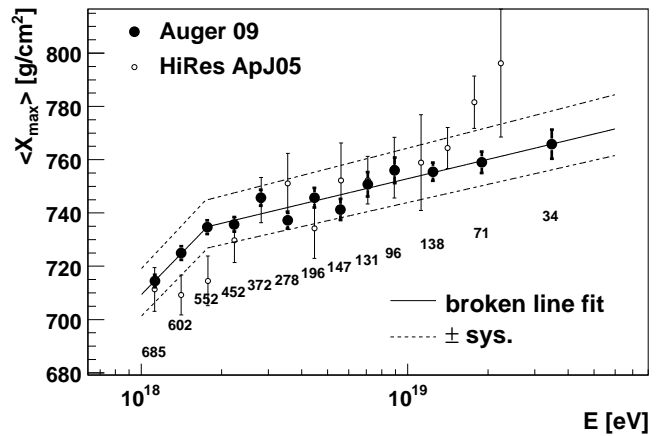}
\caption{\meanXmax as a function of energy. Lines denote a fit with
  a broken line in $\lg E$. The systematic uncertainties of
  \meanXmax are indicated by a dashed line. 
  The number of events in each energy bin is displayed
  below the data points.  HiRes data~\cite{bib:Abbasi2005ni} are shown
  for comparison.}
\label{fig_meanXmax}
\end{figure}

\begin{figure*}[!t]
\centering
\includegraphics[width=\textwidth]{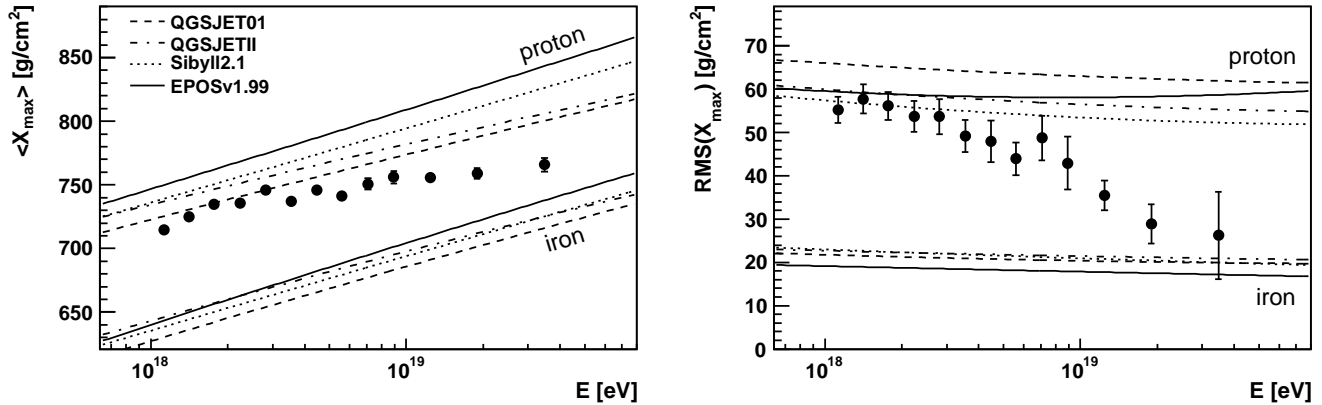}
\caption{\meanXmax and \sigmaXmax  compared
         with air shower simulations~\cite{bib:conex} using different 
         hadronic interaction models\cite{bib:simulations}.}
\label{fig_meanAndRMS}
\end{figure*}

{\it Results and Discussion.} -- The measured \meanXmax and \sigmaXmax
values are shown in Figs.~\ref{fig_meanXmax} and~\ref{fig_meanAndRMS}.
We use bins of $\Delta\lg E=0.1$ below 10~\EeV{} and $\Delta\lg E=0.2$
above that energy.  The last bin starts at \energy{19.4}, integrating
up to the highest energy event ($E=(59\pm8)$~\EeV).  The systematic
uncertainty of the FD energy scale is 22\%~\cite{bib:icrcflux}.
Uncertainties of the calibration, atmospheric conditions,
reconstruction and event selection give rise to a systematic
uncertainty of $\le$\depth{13} for $\meanXmax$ and $\le$\depth{6} for
the RMS.  The results were found to be independent of zenith angle,
time periods and FD stations within the experimental uncertainties.

A fit of the measured \meanXmax values
 with a constant elongation rate does not describe our data
($\chi^2$/Ndf=34.9/11), but as can be seen in Fig.~\ref{fig_meanXmax},
using two slopes yields a satisfactory fit ($\chi^2$/Ndf=9.7/9)
with an elongation rate of \lowD below \breakE and \highD above this
energy. If the properties of hadronic interactions do not change
significantly over less than two orders of magnitude in primary energy
($<$ factor 10 in center of mass energy), this change of $\Delta
D_{10}=$\unit[$(82^{+35}_{-21})$]{\gcm/decade} would imply a change in
the energy dependence of the composition around the ankle, supporting
the hypothesis of a transition from galactic to extragalactic cosmic
rays in this region.

The \meanXmax result of this analysis is compared to the HiRes
data~\cite{bib:Abbasi2005ni} in Fig.~\ref{fig_meanXmax}. Both
data-sets agree well within the quoted systematic uncertainties. The
$\chi^2/\mathrm{Ndf}$ of the HiRes data with respect to the broken-line 
fit described above is $20.5/14$.  This value reduces
to $16.8/14$ if a relative energy shift of 15\% is applied, such as
suggested by a comparison of the Auger and HiRes energy
spectra~\cite{bib:gzkmeas}.

The shower-to-shower fluctuations, \sigmaXmax, are obtained by
subtracting the detector resolution in quadrature from the width of
the observed \Xmax distributions resulting in a correction of
$\le$\depth{6}. As can be seen in the right panel of
Fig.~\ref{fig_meanAndRMS}, we observe a decrease in the
fluctuations with energy from about \highRMS~to \depth{\lowRMS}
as the energy increases.  Assuming again that the hadronic interaction
properties do not change much within the observed energy range, these
decreasing fluctuations are an independent signature of an increasing
average mass of the primary particles.

For the interpretation of the absolute values of \meanXmax and
\sigmaXmax a comparison to air shower simulations is needed.  As can
be seen in Fig.~\ref{fig_meanAndRMS}, there are considerable
differences between the results of calculations using different
hadronic interaction models. These differences are not necessarily
exhaustive, since the hadronic interaction models do not cover the
full range of possible extrapolations of low energy accelerator
data. If, however, these models provide a realistic description of
hadronic interactions at ultra high energies, the comparison of the
data and simulations leads to the same conclusions as above, namely a
gradual increase of the average mass of cosmic rays with energy up to
59 \EeV.


{\it Acknowledgments.} --
The successful installation and commissioning of the Pierre Auger Observatory
would not have been possible without the strong commitment and effort
from the technical and administrative staff in Malarg\"ue.
We are very grateful to the following agencies and organizations for financial support: 
Comisi\'on Nacional de Energ\'ia At\'omica, 
Fundaci\'on Antorchas,
Gobierno De La Provincia de Mendoza, 
Municipalidad de Malarg\"ue,
NDM Holdings and Valle Las Le\~nas, in gratitude for their continuing
cooperation over land access, Argentina; 
the Australian Research Council;
Conselho Nacional de Desenvolvimento Cient\'ifico e Tecnol\'ogico (CNPq),
Financiadora de Estudos e Projetos (FINEP),
Funda\c{c}\~ao de Amparo \`a Pesquisa do Estado de Rio de Janeiro (FAPERJ),
Funda\c{c}\~ao de Amparo \`a Pesquisa do Estado de S\~ao Paulo (FAPESP),
Minist\'erio de Ci\^{e}ncia e Tecnologia (MCT), Brazil;
AVCR AV0Z10100502 and AV0Z10100522,
GAAV KJB300100801 and KJB100100904,
MSMT-CR LA08016, LC527, 1M06002, and MSM0021620859, Czech Republic;
Centre de Calcul IN2P3/CNRS, 
Centre National de la Recherche Scientifique (CNRS),
Conseil R\'egional Ile-de-France,
D\'epartement  Physique Nucl\'eaire et Corpusculaire (PNC-IN2P3/CNRS),
D\'epartement Sciences de l'Univers (SDU-INSU/CNRS), France;
Bundesministerium f\"ur Bildung und Forschung (BMBF),
Deutsche Forschungsgemeinschaft (DFG),
Finanzministerium Baden-W\"urttemberg,
Helmholtz-Gemeinschaft Deutscher Forschungszentren (HGF),
Ministerium f\"ur Wissenschaft und Forschung, Nordrhein-Westfalen,
Ministerium f\"ur Wissenschaft, Forschung und Kunst, Baden-W\"urttemberg, Germany; 
Istituto Nazionale di Fisica Nucleare (INFN),
Ministero dell'Istruzione, dell'Universit\`a e della Ricerca (MIUR), Italy;
Consejo Nacional de Ciencia y Tecnolog\'ia (CONACYT), Mexico;
Ministerie van Onderwijs, Cultuur en Wetenschap,
Nederlandse Organisatie voor Wetenschappelijk Onderzoek (NWO),
Stichting voor Fundamenteel Onderzoek der Materie (FOM), Netherlands;
Ministry of Science and Higher Education,
Grant Nos. 1 P03 D 014 30, N202 090 31/0623, and PAP/218/2006, Poland;
Funda\c{c}\~ao para a Ci\^{e}ncia e a Tecnologia, Portugal;
Ministry for Higher Education, Science, and Technology,
Slovenian Research Agency, Slovenia;
Comunidad de Madrid, 
Consejer\'ia de Educaci\'on de la Comunidad de Castilla La Mancha, 
FEDER funds, 
Ministerio de Ciencia e Innovaci\'on,
Xunta de Galicia, Spain;
Science and Technology Facilities Council, United Kingdom;
Department of Energy, Contract Nos. DE-AC02-07CH11359, DE-FR02-04ER41300,
National Science Foundation, Grant No. 0450696,
The Grainger Foundation USA; 
ALFA-EC / HELEN,
European Union 6th Framework Program,
Grant No. MEIF-CT-2005-025057, 
European Union 7th Framework Program, Grant No. PIEF-GA-2008-220240,
and UNESCO.


\end{document}